\documentclass[10pt,letterpaper]{article}
\usepackage[top=0.85in,left=2.75in,footskip=0.75in]{geometry}

\usepackage{amsmath,amssymb}

\usepackage{changepage}

\usepackage[utf8x]{inputenc}

\usepackage{textcomp,marvosym}

\usepackage{cite}

\usepackage{nameref,hyperref}

\usepackage[right]{lineno}

\usepackage{microtype}
\DisableLigatures[f]{encoding = *, family = * }

\usepackage{booktabs}

\usepackage[table]{xcolor}

\usepackage{array}

\usepackage{subcaption}

\newcolumntype{+}{!{\vrule width 2pt}}

\newlength\savedwidth

\raggedright
\usepackage[aboveskip=1pt,labelfont=bf,labelsep=period,justification=raggedright,singlelinecheck=off]{caption}

\makeatletter
\renewcommand{\@biblabel}[1]{\quad#1.}
\makeatother

\usepackage{lastpage,fancyhdr,graphicx}
\usepackage{epstopdf}
\fancyheadoffset[L]{2.25in}
\fancyfootoffset[L]{2.25in}
\begin{document}
\vspace*{0.2in}

\begin{flushleft}
{\Large
\textbf\newline{Assessing dengue fever risk in Costa Rica by using climate variables and machine learning techniques} 
}
\newline
\\

Luis A. Barboza \textsuperscript{1\Yinyang*},
Shu-Wei Chou   \textsuperscript{2\Yinyang},
Paola V\'asquez\textsuperscript{3\textcurrency},
Yury E. Garc\'ia\textsuperscript{3,4},
Juan G. Calvo   \textsuperscript{1\ddag},
Hugo G. Hidalgo    \textsuperscript{5\ddag},
Fabio Sanchez   \textsuperscript{1}
\\
\bigskip
\textbf{1} Centro de Investigaci\'on en Matem\'atica Pura y Aplicada - Escuela de Matem\'atica, Universidad de Costa Rica, San Jos\'e, Costa Rica
\\
\textbf{2} Centro de Investigaci\'on en Matem\'atica Pura y Aplicada - Escuela de Estadística, Universidad de Costa Rica, San Jos\'e, Costa Rica.
\\
\textbf{3} Centro de Investigaci\'on en Matem\'atica Pura y Aplicada, Universidad de Costa Rica, San Jos\'e, Costa Rica.
\\
\textbf{4} Department of Public Health Sciences, University of California Davis, CA, USA.
\\
\textbf{5} Centro de Investigaciones Geofísicas and Escuela de Física, Universidad de Costa Rica, San Jos\'e, Costa Rica.
\bigskip

%
%
\Yinyang These authors contributed equally to this work.

\ddag These authors also contributed equally to this work.

\textcurrency Current Address: Dept/Program/Center, Institution Name, City, State, Country 

\dag Deceased

\textpilcrow Membership list can be found in the Acknowledgments section.

* luisalberto.barboza@ucr.ac.cr 

\end{flushleft}
\section*{Abstract}
Dengue fever is a vector-borne disease mostly endemic to tropical and subtropical countries that affect millions every year and is considered a significant burden for public health. Its geographic distribution makes it highly sensitive to climate conditions. Here, we explore the effect of climate variables using the Generalized Additive Model for location, scale, and shape (GAMLSS) and Random Forest (RF) machine learning algorithms. Using the reported number of dengue cases, we obtained reliable predictions. The uncertainty of the predictions was also measured. These predictions will serve as input to health officials to further improve and optimize the allocation of resources prior to dengue outbreaks.

\section*{Author summary}
Dengue fever is a vector-borne disease that is endemic to tropical and subtropical countries. It is transmitted by mosquitoes and affects approximately 100 million people every year. Despite the fact that most infections are mild or asymptomatic, some may cause severe disease known as dengue shock syndrome or dengue hemorrhagic fever. Using statistical tools to provide insight to public health officials regarding the risk of dengue in different cantons of the country gives them the necessary time to prepare and better allocate resources.


\section*{Introduction}

Dengue virus transmission represents a public health challenge for countries in tropical and subtropical regions worldwide~\cite{brady2012refining}. For the past decades, the increasing geographical spread of the pathogen and its two main vectors, \textit{Aedes aegypti} and \textit{Aedes albopictus}~\cite{messina2014global}, has led to the development and implementation of multiple prevention and control measures~\cite{gubler1998dengue}. However, the complex interactions and constant variations in population mobility, socioeconomic, demographic, environmental, and climate factors that modulate the spatial and temporal distribution of the disease make it difficult to achieve timely, effective, and sustainable strategies. 

In this context, researchers around the world are increasingly working towards the development of innovative, tailored, and cost-effective tools that enhance the design of public health policies for vector-borne diseases founded upon the rapid systematization and analysis of information, as well as an increase in interdisciplinary collaboration~\cite{world2012global}. In these efforts, statistical and machine learning techniques are increasingly used for public health surveillance and epidemiological modeling~\cite{wiemken2019machine}. These methods efficiently estimate and predict health outcomes from epidemiological, entomological, and other climate and social determinants surveillance. Through computational algorithms, this branch of artificial intelligence facilitates the possibility of integrating scientific knowledge, processing large databases, learning from past documented reported cases, and ultimately projecting transmission tendencies to identify and target the most vulnerable at-risk areas. Dengue is a climate-sensitive disease, where changes in temperature, humidity, and precipitation affect the biology, behavior, and availability of the mosquito to reproduce, develop, propagate the virus, and interact with the human host~\cite{ebi2016dengue,naish2014climate,tran2020estimating}. The use of satellite imagery and weather monitoring as input data in machine learning models and other statistical learning approaches have shown promising results~\cite{lowe2011spatio,lowe2013development} that could effectively predict the relative risk of dengue transmission.

In Costa Rica, a country of 5,213,362 inhabitants~\cite{Bienveni32:online}, administratively divided into seven provinces and 82 cantons, there has been endemic circulation of three of the four dengue virus serotypes (DENV-1, DENV-2, DENV-3) and more than 393,000 reported cases by the Ministry of Health since 1993~\cite{SitioWeb58:online}. The country has a tropical climate, which provides ideal conditions for the mosquito vector to survive, replicate and transmit disease. The interaction between geographic, oceanic, and atmospheric factors divides Costa Rica into seven major climatic regions~\cite{ClimaenC13:online}, each further subdivided into sub-regions. Micro-climates separated by short distances make it crucial to customize dengue transmission risk analysis in a country where health authorities do not formally use weather information as input in developing prevention and control activities. 


In this article, we showed the results of using two different modeling approaches, GAMLSS and RF, to forecast the relative risk of dengue infections in 32 municipalities of Costa Rica. The analysis is a continuation of previous work in~\cite{vazquez2020}, where an initial approach using Generalized Additive Models and RF allowed us to retrospectively predict the relative risk of dengue for 2017 in five diverse climate municipalities, using as input the information of five weather stations provided by the National Meteorological Institute~\cite{vazquez2020}. Based on those results, an initial collaboration with Costa Rican health authorities allowed identifying 32 municipalities of interest based on their entomological and epidemiological characteristics. This work also incorporates satellite information to produce monthly dengue forecasts. We then evaluated the forecast with available dengue reports from the Ministry of Health.


\section*{Data description}

\begin{figure}
	\centering
	\includegraphics[scale=0.8]{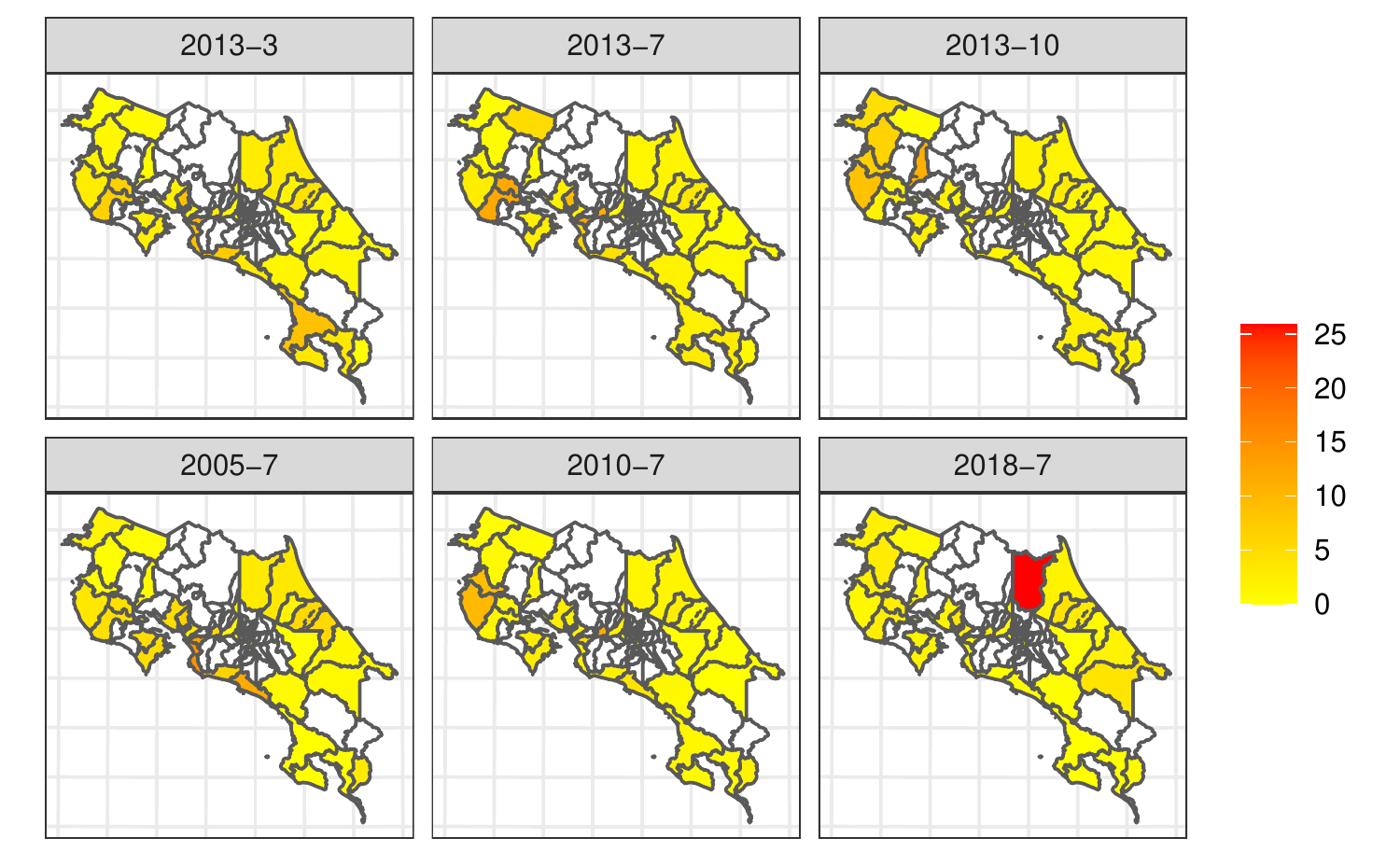}
	\caption{Relative Risk ($RR$) over the 32 cantons under study for different months and years of available data.}
	\label{RRplot}
\end{figure}

The predictive models applied in this article used the following sources of information:

\begin{description}
	\item[Dengue Data:] Data of clinically suspected and confirmed monthly cases of dengue fever in Costa Rica. The data is collected from all the local administrative areas (municipalities or cantons) in the country, covering the years 2000-2021, and provided by the Ministry of Health. The Ministry of Health has a particular interest in the monthly forecast of 32 specific cantons due to their particular epidemiological behavior. In order to quantify the relative incidence of dengue cases at the $i$-th canton compared with the incidence in the country at time $t$ (monthly basis), we use the relative risk ($RR$):
	\begin{align*}
		RR_{i,t}=\frac{\frac{\text{Cases}_{i,t}}{\text{Population}_{i,t}}}{\frac{\text{Cases}_{CR,t}}{\text{Population}_{CR,t}}}
	\end{align*}  
    where $\text{Cases}_{i,t}$ ($\text{Population}_{i,t}$)  and $\text{Cases}_{CR,t}$ ($\text{Population}_{CR,t}$) are the number of observed dengue cases (population size) at canton $i$ and country-level respectively, at time $t$. The overall behavior of the relative risks, at three specific months in 2013 (first row), and three specific years in July (second row) is shown in Figure~\ref{RRplot}). Most of the studied cantons are located near the Pacific and Caribbean Oceans, and there is inter-monthly and interannual variation in the behavior of the relative risk.
    
	\item[Climate Data:] the covariate information consists of:
	\begin{itemize}
		\item Daily Precipitation estimates ($P_{i,t}$) were used to index land surface rainfall. Data was obtained from the Climate Hazards Group InfraRed Precipitation with Station data (CHIRPS); see~\cite{FunkChris2015Tchi}. Due to the spatial high-resolution nature of this dataset (\~5km by 5 km), we were able to compute monthly cumulative rainfall estimates for each canton by adding the same estimate over smaller administrative areas (\textit{distritos}).   
		\item Weekly ENSO Sea Surface Temperature Anomaly ($S_{i,t}$), also known as SSTA index. Data was obtained from the Climate Prediction Center (CPC) of the United States National Oceanographic and Atmospheric Administration (NOAA).
		
		\item Normalized Difference Vegetation Index (NVDI) ($N_{i,t}$), an index of the greenness of vegetation for a 16-day time resolution and 250m spatial resolution. It was obtained from the Moderate Resolution Imaging Spectroradiometer (MODIS) satellite, and available through the \verb|MODISTools| R package (see~\cite{TuckPhillips})  
		\item Daytime Land Surface Temperature ($L_{i,t}$) in Kelvin degrees for a 8-day time resolution and 1km spatial resolution. Obtained using the same resources as the NDVI covariate.
		\item Tropical Northern Atlantic Index ($TN_{i,t}$). Anomaly index of the sea-surface temperature over the eastern tropical North Atlantic Ocean (see~\cite{EnfieldDavidB1999Huit}). This index is used, because previous work in the region, such as \cite{Hidalgo2017-hr}, suggested that the inclusion of SST information from the Caribbean/Atlantic improves performance compared to forecasts produced with only Pacific Ocean ENSO conditions.   
	\end{itemize}
\end{description}

\section*{Methods}

\subsection*{Fitting Stage}

As a first step to model the relationship among the climate covariates and the relative risk of dengue fever at a fixed canton, we incorporate the historical delayed associations between those variables by applying a Distributed Lag Non-Linear Model (DLNM) framework~\cite{Gasparrini2010,Gasparrini2014}. It consists of a bi-dimensional space of functions that specifies a exposure-lag-response function $f \cdot w (x,l)$ which depends on the predictor $x$ along the time lags $l$ in a combined way. This combination allows the possibility of specifying a non-linear and delayed association between the climate covariate and the dengue incidence. For each covariate we take into account a maximum exposure of 18 months in its lag representation, based on the cross-correlation and wavelet behavior among the series (see \cite{garcia2021wavelet}) and a b-spline or linear basis representation on the variable space. The procedure was performed with the R package \verb|dlnm|~\cite{Gasparrini2011}. 

The structure of the models analyzed in this article in terms of dependent variable and covariates for a fixed location is as follows:
\begin{align}
	RR_t \sim f(RR_{t-1},C_1P_t, C_2S_t, C_3N_t, C_4L_t, C_5TN_t,M_t)	
	\label{estructura}
\end{align}
where $f$ is a function depending on the method employed, the matrices $C_i$ are defined in terms of the DLNM representation and $M_t$ is a factor-type variable describing the monthly fixed effect. Recall that the time $t$ has month as unit. 

The first method that we use for $f$ is the generalized additive model for location, scale and shape (\textbf{GAMLSS}). It represents a generalization of the GAM method used in~\cite{vazquez2020}. It is a flexible class of statistical framework where the location, scale, skewness and kurtosis parameters from the distribution of the response variable can be modeled as a additive function of covariates~\cite{stasinopoulos2017}. In this particular case the model can be written as:
\begin{align}
	RR_t &\overset{ind}{\sim} \mathcal{D}(\mu,\sigma,\nu) \notag\\
	g_1(\mu)&= \beta_{10}+\beta_{11}RR_{t-1}+\beta_{12}C_1P_t+ \beta_{13}C_2S_t+\beta_{14}C_3N_t+\beta_{15}C_4L_t+\beta_{16}C_5TN_t+\beta_{17} M_t \notag \\
	g_2(\sigma)&=\beta_{20}\\
	g_3(\nu)&=\beta_{30} \notag
\end{align}
where the response variable is distributed as a three-parameter distribution $\mathcal{D}$, $\mu$ is the location, $\sigma$ is the scale, $\nu$ is the parameter related to the skewness of the distribution and $g_i$ for $i=1,2,3$ are link functions.

Due to the fact that the relative risk of dengue is a non-negative skewed variable with a large frequency of zeros ($16.2\%$), mixed distributions which have positive domain with a positive probability at zero are appropriate choices for modeling purposes. In this case, the zero adjusted gamma distribution (ZAGA) and the zero adjusted inverse Gaussian (ZAIG) are considered. Results for both choices are similar, and therefore we show our results for the ZAGA distribution only.

The ZAGA density function is defined by the mixed continuous-discrete probability density:

\begin{equation*}
f_Y(y) =\left\{
\begin{array}{ll}
\nu \quad \text{~~~~~~~~~~~~~~~if } y=0 \\
(1-\nu)f_W(y)  \text{~~if~~} 0<y<\infty
\end{array}
\right.
\end{equation*}
for $0 \leq y < \infty$, where $W \sim GA(\mu,\sigma)$ is a gamma distribution with $0<\mu<\infty$, $0<\sigma<\infty$ and $0<\nu<1$, i.e.
\begin{align*}
f_W(y|\mu,\sigma)=\frac{1}{(\sigma^2\mu)^{1/\sigma^2}} \frac{y^{\frac{1}{\sigma^2}-1} e^{-y/(\sigma^2\mu)}}{\Gamma(1/\sigma^2)},	
\end{align*}
for $y>0$, $\mu>0$ and $\sigma>0$. The advantage of this parametrization is that $E(W)=\mu$ and $V(W)=\sigma^2\mu^2$. For the GAMLSS specification, $ZAGA(\mu,\sigma,\nu)$ defines the log link functions for $\mu$ and $\sigma$, i.e., $g_1(\mu)=\log(\mu)$ and $g_2(\sigma)=\log \sigma$; and the logit link function for $\nu$, i.e. $g_3(\nu)=\log \left[ \nu/(1-\nu) \right]$. 

The second method uses a Random Forest (\textbf{RF}) approach. This method is based on the construction of bootstrapped ensemble of regression trees and combined such that the prediction variance can be reduced (see~\cite{BreimanLeo2001RF} and~\cite{Hastie2009}). One of the main advantages of this method is the reduced number of tuning parameters that eases its computational manipulation and stability~\cite{BreimanLeo2001RF}. 	

The fitting process of the GAMLSS and RF models was performed with the R packages \verb|gamlss|~\cite{rigby2005} and \verb|ranger|~\cite{Wright2017}, respectively.

\subsection*{Prediction Stage}

Equation~\eqref{estructura} has been fitted over a certain calibration period using any of the two methodology alternatives, we forecast the relative risk over a testing period using the past information of climatic covariates and the relative risk itself. Provided that the response variable in~\eqref{estructura} depends on the current values of the climatic covariates, it is crucial to select an appropriate method to obtain the climate predictions in the near future, so that it can supply accurate inputs to our predictive model under both methodologies. Since the climate covariates used in this study are highly correlated, a suitable method to describe and predict their interaction is the vector auto-regressive (VAR) model (see more details in~\cite{Tsay2015}). For each canton, we include the trend and seasonal factors to fit a VAR model and select the best lag order based on BIC criterion, over the training period (which is the same period as the one used for the fitting~\eqref{estructura}). Then, we jointly forecast over the testing period the covariates. Figure \ref{climate-VAR} shows an illustration of the observed climate covariates along with its forecast values at Alajuela and Quepos cantons. These predictions together with the predicted relative risks are used in order to provide forecasts of the dependent variable over the testing period. Finally, in order to assess the prediction uncertainty, we apply a non-parametric bootstrap~\cite{Efron1993} and construct prediction intervals using the corresponding forecasts for each bootstrap step.

\subsection*{Model Comparison}

We use two different metrics to compare the predictive performance of each methodology for each fixed location. The normalized Mean-Squared Error ($NRMSE$):
\begin{align*}
	NRMSE = \sqrt{\frac{1}{m\overline{ RR}}\sum_{t=1}^m(RR_t-\widehat {RR}_t)^2}
\end{align*}
where $m$ is the number of months in the testing period, $\overline{RR}$ is the mean relative risk over the same period and $\widehat{RR}$ is the estimated relative risk according to any of the two models. The normalized Interval Score at $\alpha$ level ($NIS_{\alpha}$) is the normalized version of the Interval Score (see~\cite{GoodProbabilityAssessors} and~\cite{Gneiting2007a}) so we can compare different locations regardless the scale of their corresponding relative risk:
\begin{align*}
	NIS_{\alpha}=\frac{1}{m\bar RR} \sum_{t=1}^m\left[(U_t-L_t)+\frac{2}{1-\alpha}(L_t-RR_t)\cdot 1_{RR_t<L_t}+\frac{2}{1-\alpha}(RR_t-U_t)\cdot 1_{RR_t>U_t}\right]
\end{align*}
The latter metric is more complete than the former in the sense of evaluating the predictive capacity of the models when the uncertainty is summarized through an predictive interval~\cite{Gneiting2007a}.  

\section*{Results}

We used the dengue and climate data as described above to fit the model in~\eqref{estructura} using both the GAMLSS and RF methodologies. The training period includes monthly observations for the 32 \textit{cantons} in the study, from January 2000 to December 2020. This period was considered related to available satellite data and epidemiological information. The DLNM basis was chosen to be linear in the variable and lag space for the TNA index, Land Surface Temperature, and NDVI, whereas a B-spline basis is assumed for the variable space, and it is linear for the lag space for precipitation and ENSO. These choices allow an acceptable balance among the complexity of the models and predictive precision over all the locations.     

\begin{figure}
	\centering
	\includegraphics[scale=0.2]{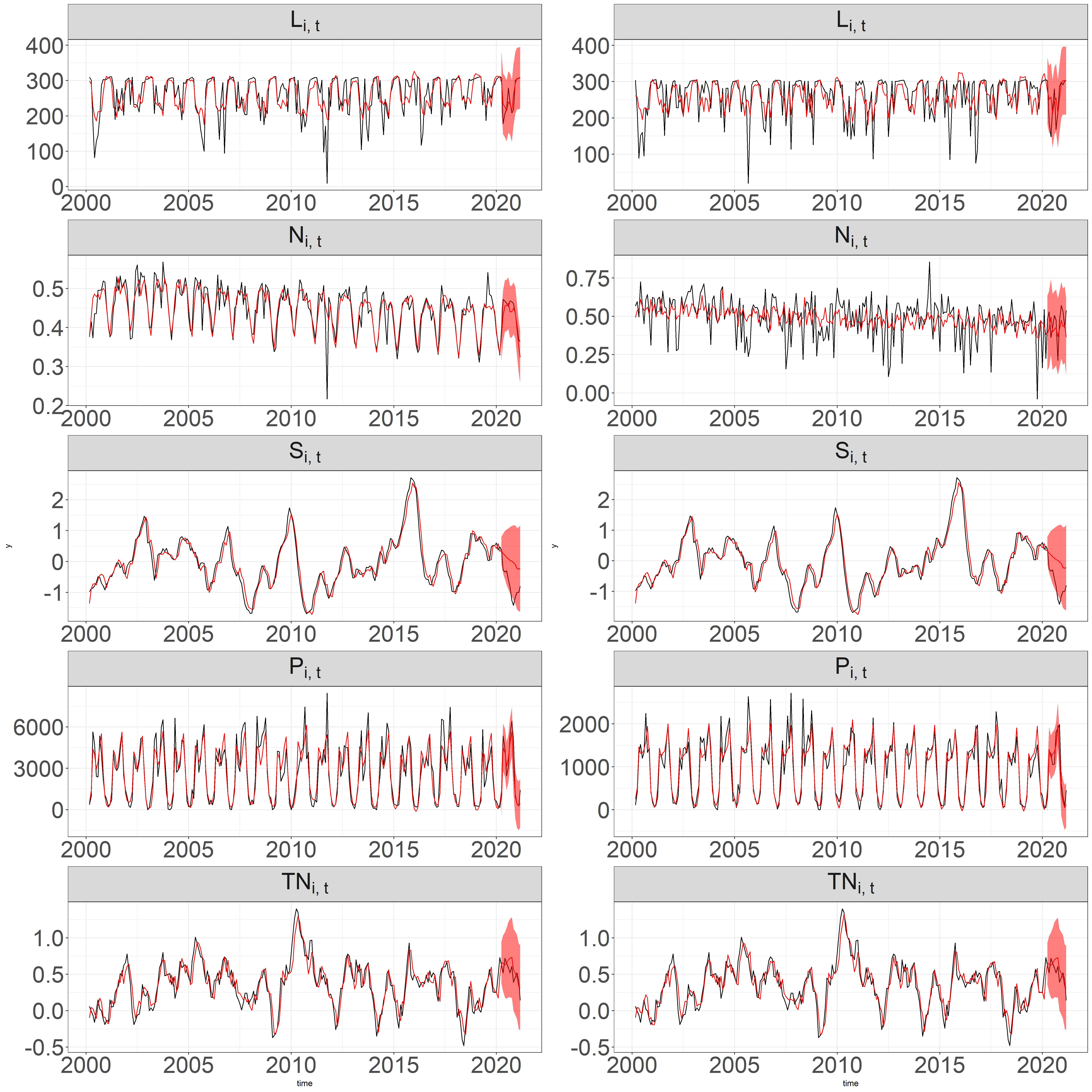}
	\caption{Observed climate covariates and forecast values at two specific cantons: Alajuela (left panels) and Quepos (right panels). Black line: observed climate covariates, red line: forecast values, and red shaded areas: $95\%$ confidence regions.}
	\label{climate-VAR}
\end{figure}

\begin{figure}
	\centering
	\includegraphics[scale=0.65]{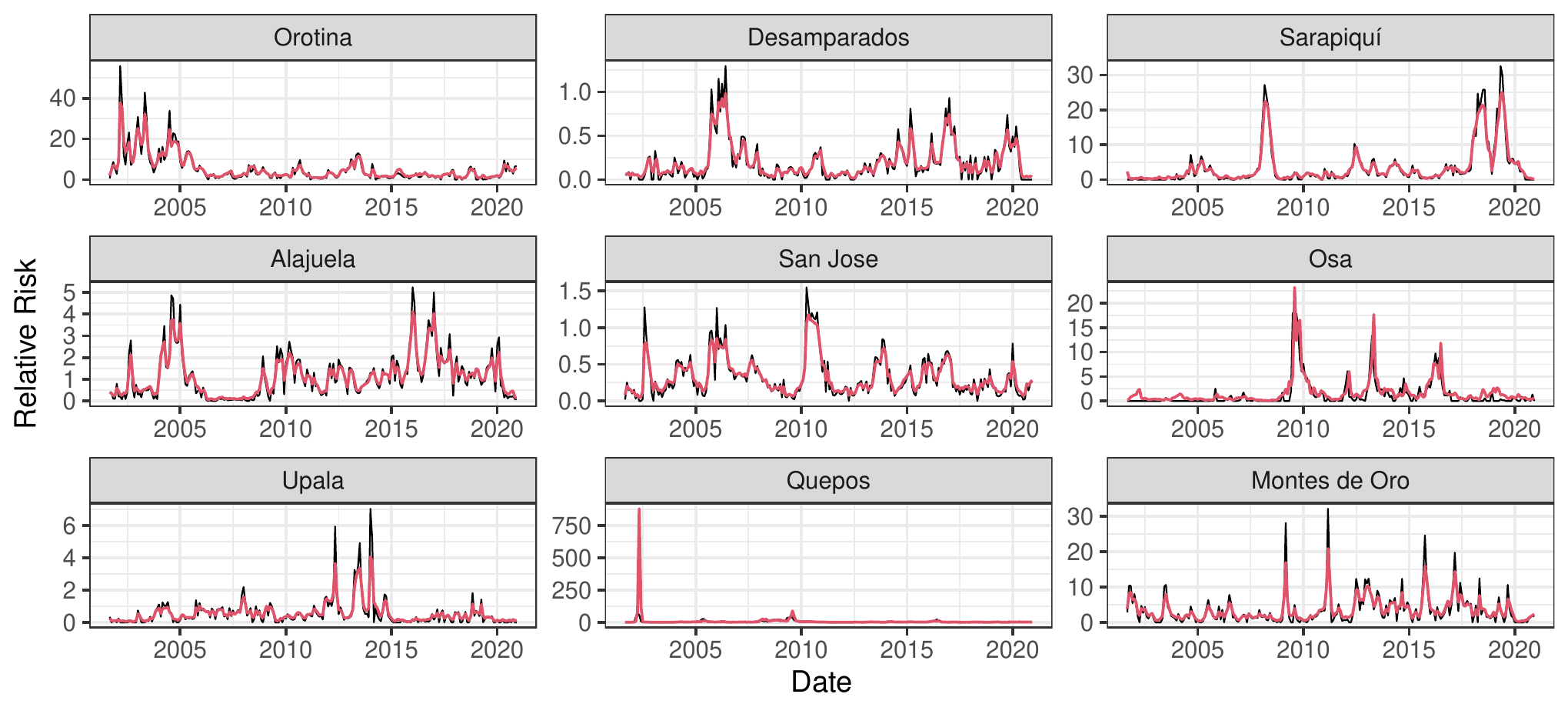}
	\caption{Comparison over the fitting period. Upper six panels: best cantons according to NIS metric. Lower three panels: worst cantons according to NIS metric. Black line: observed $RR$, red line: estimated $RR$.}
	\label{training-plot}
\end{figure}

\begin{figure}
	\centering
	\includegraphics[scale=0.65]{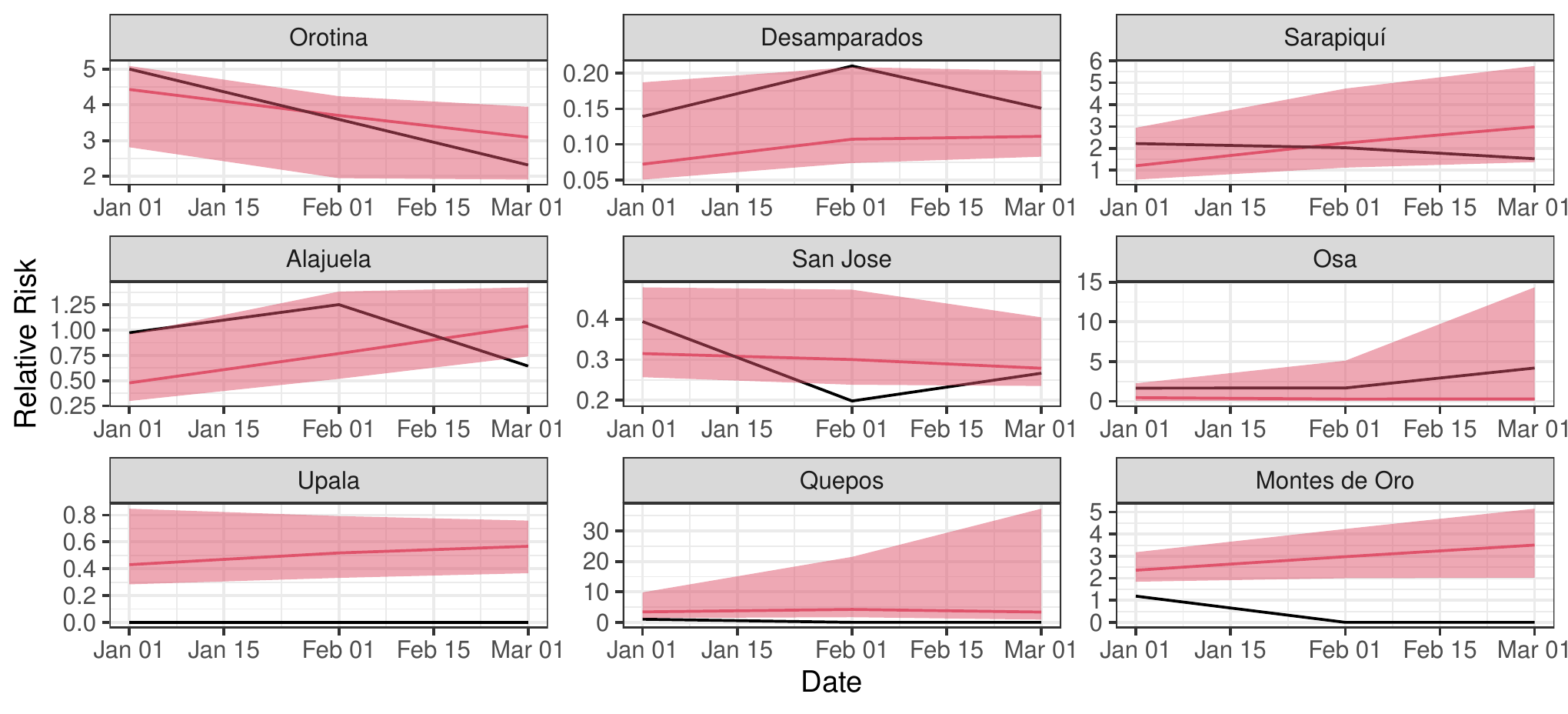}
\caption{Forecast comparison over the testing period. Upper six panels: best cantons according to NIS metric. Lower three panels: worst cantons according to NIS metric.}
\label{testing-plot}
\end{figure}

Once the transformed covariates in~\eqref{estructura} are determined, we fitted both methodologies over the 32 locations individually. Then adjusted a VAR model using the climate information for each location over the training period to obtain predicted values of the covariates. Hence, we can predict the relative risk of dengue for the first three months of 2021. Figure~\ref{climate-VAR} shows the fitted values and predictions of the climate covariates for Alajuela and Quepos. We observed that the general features like trend and seasonality of the multiple time series are well-captured for the testing period.  
Due to the auto-regressive nature of the model in~\eqref{estructura}, we used the predicted value of $RR_t$ as a covariate in the prediction of $RR_{t+1}$. Once the predicted relative risks over the first three months of 2021 are computed, we compare the observed and the predicted values with the $NRMSE$ and $NIS_{.95}$ metrics and show the behavior of the best six \textit{cantons} and worst three \textit{cantons} according to the latter metric, regardless of the method employed. The comparison is shown for the training period in Figure~\ref{training-plot}    
and the testing period in Figure \ref{testing-plot}. Moreover, there is not a significant difference among the methods in terms of the $NIS_{95}$ metric, as can be seen in \nameref{S1_Fig}. \nameref{S1_Table} shows which model is chosen for each location and their respective metrics, where in general, there is not a model that is predominant over all the locations.

Together with the expected values of the relative risks in Figure~\ref{testing-plot}, we computed predicted uncertainties at $95\%$-level using a blocked non-parametric bootstrap~\cite{Efron1993} with 100 replicates and a block size of six months, over the testing period only.
	
In general terms, the fitting performs well in the training period, except for some extreme observations that the model does not capture closely, for example, in the cantons of Quepos and Osa. The monthly trend over the testing period is well captured in cantons like Orotina and Desamparados, and it is partially captured as Alajuela and San Jose. The uncertainty is wide enough to conserve the trend information while it covers most of the observed values throughout the best six cantons. The worst-behaved cantons do not necessarily have a significant uncertainty compared with the well-behaved ones, while the bias is present in all the cases, where perhaps under-reporting of observed cases can be a possible cause of such difference. Forecast comparison over the testing period. Upper six panels: best cantons according to NIS metric. Lower three panels: worst cantons according to NIS metric.


\section*{Discussion}

Climate conditions in Costa Rica make it ideal for the proliferation of mosquitoes and, simultaneously, a challenging territory for the epidemiological surveillance of dengue. The importance of climate information regarding the incidence of dengue fever has been well established in previous studies~\cite{naish2014climate,ebi2016dengue}.

The understanding of the ecology of the \textit{Ae. aegypti} mosquito for the development of climate-based early warning systems for dengue prediction is crucial. In Costa Rica, dengue dynamics change geographically and temporally due to microclimates. Traditionally, health officials carry out localized studies and explore different methodologies to determine what works well depending on geographic location. Furthermore, this is an opportunity for health authorities to have a tool that provides situational insight into the risk of dengue infection in the country.



Here, we implemented a Generalized Additive Model and a Random Forest Model using mean temperature, precipitation, relative humidity, and weekly ENSO SSTA to predict the relative risk of dengue in five of the 32 riskiest cantons in this study. Although models showed a successful performance in the predictions, their efficacy and capacity were limited by the availability and quality of meteorological data. In addition, knowing the climatic covariates was a necessary condition to predict dengue behavior. In this article, these limitations were overcome using satellite information and proposing a VAR model to obtain climate predictions, which are incorporated in the models to forecast the relative risk.  

Models were trained using data from 2000 to 2020. Results showed that the VAR model performs well-predicting climatic variables, letting prospective analyses be carried out instead of retrospective studies. The GAMLSS and the Random Forest model successfully predict relative dengue risk in the first three months of 2021, corresponding to the testing period. The GAMLSS flexibility allows capturing the dynamic in cantons with a low number of cases. 

Dengue outbreaks affect the health system and the economy of a country. A person with mild symptoms of dengue can be disabled on average for seven days~\cite{de2013guide}, which generates a reduction in the workforce affecting the productivity of companies. The costs of the Costa Rican Social Security Fund in care for users with dengue were estimated at \$20.3 million for 2013, which includes care for hospitalized patients, medical consultations, and disabilities. The Ministry of Health has estimated at \$6.5 million the investment in preventive campaigns and combat actions in this same period~\cite{PGR}.

Dengue disease surveillance and vector surveillance are supposed to detect outbreaks early, reducing the social and economic impact. The cost of preventive measures has been reported to be less than the cost of treating an outbreak~\cite{stahl2013cost}. Therefore, early prediction tools are valuable because they allow taking preventive measures and directing and optimizing resources, particularly in countries like Costa Rica, where economic resources are limited.

The predictive capacity of machine learning models and the greater use of non-traditional information in the development of monitoring efforts allows more comprehensive surveillance and early identification of possible outbreaks, as well as opens the possibility to diversify the inclusion of the community in the identification of priority areas in collaboration with local governments and authorities. 

Public health authorities must be prepared to face the future challenges that vector-borne diseases pose. External factors such as climate change and the emergence of new infectious diseases highlight the need for a preventive approach instead of a reactive one as the basis for the formulation of health strategies in the country. 




\section*{Supporting information}

\paragraph*{S1 Fig.}
\label{S1_Fig}
{\bf Comparison of the distribution of NIS metric among methods.} 
\begin{figure}[htp]
	\centering
	\includegraphics[scale=0.5]{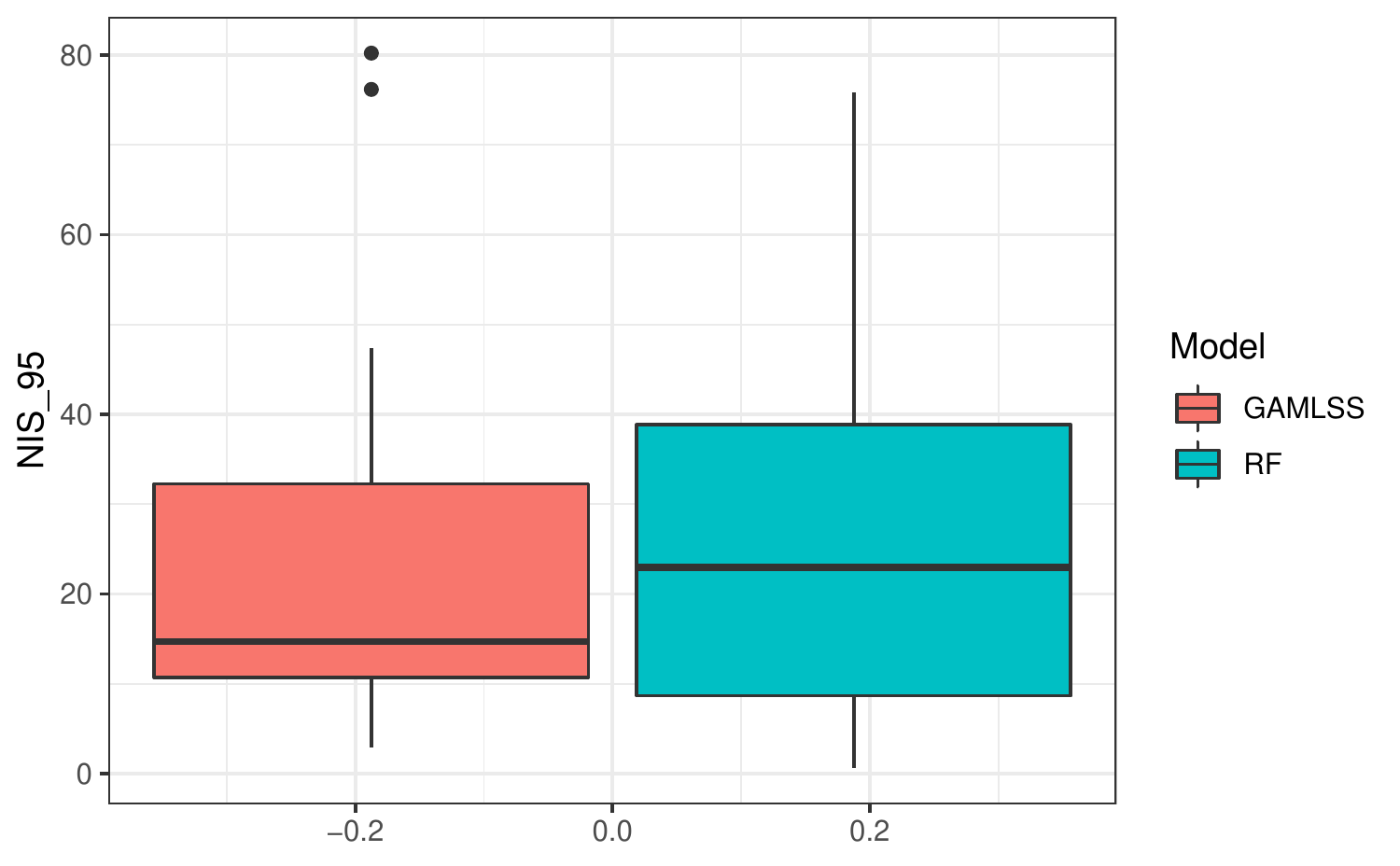}
\end{figure}


%
%
%

\paragraph*{S1 Table.}
\label{S1_Table}
{\bf Best model for each \textit{canton}} 
\begin{table}[htp!]
	\begin{tabular}[t]{lrrl}
		\toprule
		\textbf{Canton} & $NRMSE$ & $NIS_{95}$ & \textbf{Best Model}\\
		\midrule
		Alajuela & 0.2218 & 2.2789 & RF\\
		Alajuelita & 0.2650 & 13.9639 & GAMLSS\\
		Atenas & 1.7615 & 16.6668 & GAMLSS\\
		Cañas & 0.4572 & 8.8197 & GAMLSS\\
		Carrillo & 0.7468 & 13.2383 & GAMLSS\\
		Corredores & 7.4754 & 32.9063 & RF\\
		Desamparados & 0.0333 & 0.9103 & RF\\
		Esparza & 0.5269 & 13.5993 & RF\\
		Garabito & 32.7929 & 151.7930 & GAMLSS\\
		Golfito & 0.7657 & 13.0531 & RF\\
		Guacimo & 1.9182 & 8.1629 & RF\\
		La Cruz & 9.1306 & 127.2412 & RF\\
		Liberia & 2.2955 & 57.4666 & RF\\
		Limon & 2.1313 & 18.4764 & RF\\
		Matina & 3.9168 & 31.3834 & GAMLSS\\
		Montes de Oro & 18.9354 & 162.9258 & RF\\
		Nicoya & 5.3693 & 47.4291 & GAMLSS\\
		Orotina & 0.0860 & 0.6048 & RF\\
		Osa & 2.4703 & 2.8810 & GAMLSS\\
		Parrita & 8.0959 & 33.1650 & GAMLSS\\
		Perez Zeledón & 1.7181 & 14.7112 & GAMLSS\\
		Pococí & 0.6451 & 10.2410 & RF\\
		Puntarenas & 1.7821 & 22.3798 & GAMLSS\\
		Quepos & 34.3754 & 179.4324 & GAMLSS\\
		San Jose & 0.0195 & 2.5938 & RF\\
		Santa Ana & 0.6331 & 31.8002 & RF\\
		SantaCruz & 0.7159 & 13.4637 & GAMLSS\\
		Sarapiquí & 0.5572 & 1.7992 & RF\\
		Siquirres & 0.8202 & 6.6540 & RF\\
		Talamanca & 10.9225 & 26.2873 & RF\\
		Turrialba & 0.9400 & 7.0685 & GAMLSS \\
		\bottomrule
	\end{tabular}
\end{table}

\section*{Acknowledgments}

\nolinenumbers

%
%
%

\bibliography{references}

\end{document}